# The LOFT Wide Field Monitor

S. Brandt[*a], M. Hernanz[b], L. Alvarez[b], P. Azzarello[c], D. Barret[d], E. Bozzo[c], Budtz-Jørgensen[a], R. Campana[e,f], , E. del Monte[e,f], I. Donnarumma[e,f], Y. Evangelista[e,f], M. Feroci[e,f], J.L. Galvez Sanchez[b], D. Götz[g], F. Hansen[a], J.W. den Herder[h], R. Hudec[i], J. Huovelin[j], D. Karelin[b], S. Korpela[j], N. Lund[a], P. Orleanski[k], M. Pohl[l], A. Rachevski[o], A. Santangelo[m], S. Schanne[g], C. Schmid[n], S. Suchy[m], C. Tenzer[m], A. Vacchi[o], J. Wilms[n], G. Zampa[o], N. Zampa[o], J. in't Zand[h], A. Zdziarski[p],

[a]DTU Space, Elektrovej Building 327, DK-2800, Kgs. Lyngby, Denmark;
[b]IEEC/CSIC, Campus UAB, E-08193, Bellaterra, Spain;
[c]ISDC, Chemin d'Ecogia 16, CH-1290, Versoix, Switzerland;
[d]IRAP, 9, Avenue du Colonel Roche, BP 44346, 31028 Toulouse Cedex 4, France;
[e]INAF/IAPS, Via Fosso del Cavaliere 100, I-00133, Roma, Italy;
[f] INFN/Sezione di Roma 2, Via della Ricerca ScientiScientica 1, I-00133, Roma, Italy;
[g]CEA Saclay, DSM/DAPNIA/Service d'Astrophysique, 91191 Gif sur Yvette, France;
[h]SRON, Sorbonnelaan 2, 3584 CA Utrecht, The Netherlands;
[i]Astronomical Institute of the ASCR, v.v.i. and Czech Technical Univ. in Prague, Czech Republic;
[j]Department of Physics, University of Helsinki, Gustaf Hällströmin katu 2a, FI-00560, Helsinki;
[k]Space Research Centre, Polish Academy of Sciences, Bartycka 18A, Warsaw, Poland;
[l]DPNC, University of Geneva, Switzerland;
[m]IAAT, Sand 1, D-72076, Tübingen, Germany;
[n]University of Erlangen-Nuremberg & Erlangen Centre of Astroparticle Physics;
[o]INFN-Trieste, Padriciano 99, I-34127, Trieste, Italy;
[p]N. Copernicus Astronomical Center, Bartycka 18, 00-716 Warsaw, Poland;

**ABSTRACT**

LOFT (Large Observatory For x-ray Timing) is one of the four missions selected in 2011 for assessment study for the ESA M3 mission in the Cosmic Vision program, expected to be launched in 2024. The LOFT mission will carry two instruments with their prime sensitivity in the 2-30 keV range: a 10 m$^2$ class large area detector (LAD) with a <1° collimated field of view and a wide field monitor (WFM) instrument based on the coded mask principle, providing coverage of more than 1/3 of the sky. The LAD will provide an effective area ~20 times larger than any previous mission and will by timing studies be able to address fundamental questions about strong gravity in the vicinity of black holes and the equation of state of nuclear matter in neutron stars. The prime goal of the WFM will be to detect transient sources to be observed by the LAD. However, with its wide field of view and good energy resolution of <300 eV, the WFM will be an excellent monitoring instrument to study long term variability of many classes of X-ray sources. The sensitivity of the WFM will be 2.1 mCrab in a one day observation, and 270 mCrab in 3s in observations of in the crowded field of the Galactic Center. The high duty cycle of the instrument will make it an ideal detector of fast transient phenomena, like X-ray bursters, soft gamma repeaters, terrestrial gamma flashes, and not least provide unique capabilities in the study of gamma ray bursts. A dedicated burst alert system will enable the distribution to the community of ~100 gamma ray burst positions per year with a ~1 arcmin location accuracy within 30 s of the burst. This paper provides an overview of the design, configuration, and capabilities of the LOFT WFM instrument.

---

[*] sb@space.dtu.dk; phone +45 4525 9710; www.space.dtu.dk

# 1. INTRODUCTION

LOFT (Large Observatory For x-ray Timing) [1][2], is one of the four missions selected in 2011 for assessment study for the ESA M3 mission in the Cosmic Vision program [3]. LOFT will carry two science instruments; both based on Silicon drift detectors (SDDs). The Large Area Detector (LAD) is a collimated instrument with an effective area of ~10m$^2$, designed for X-ray timing [4]. Through innovative technology LAD will provide an effective area ~20 times larger than any previous mission. The spectral resolution of the LAD will be ~250 eV or better at end of life, and with a time resolution better than 10µs.

The second instrument on LOFT is a Wide Field Monitor (WFM) based on the coded mask principle, and with a detector plane of SDDs much similar to the LAD detectors, but with a design optimized for position determination. The WFM is primarily needed to detect targets of opportunity for the LAD, but as a consequence with interesting and unique capabilities on its own.

The mission duration will be 4 years, and is mainly driven by the statistics of the occurrence of the bright black hole transients, which are a class of prime targets for the LAD.

The original LOFT concept was based on a payload compatible with a Vega launch vehicle. However, during the assessment phase, it has become clear that the more powerful Soyuz launcher will be needed in order to meet its science requirements, which are not compatible with an effective area of the LAD less than ~10m$^2$. The capabilities of the Soyuz launcher in terms of mass and volume will relax several constraints and allow optimizing the spacecraft systems and the payload for maximum performance without major cost increases. One example is achieving the ideal orbit for LOFT, which is equatorial with an altitude of ~550 km in order to reduce the influence of the South Atlantic Anomaly, as radiation damage is driving the end-of-life spectral resolution of the Silicon drift detectors to be used.

The main telemetry station will be Kourou, but ASIs Malindi station will also be used in order to meet the requirements of downloading the large data volume generated by the science instruments, averaging ~6.7 Gbits or more per orbit.

The assessment phase will be concluded in 2013 and ESA is then expected to select one of the M3 candidate missions for an expected launch in 2024. However, schedules must, for ESA programmatic reasons, remain compatible with a late 2022 launch.

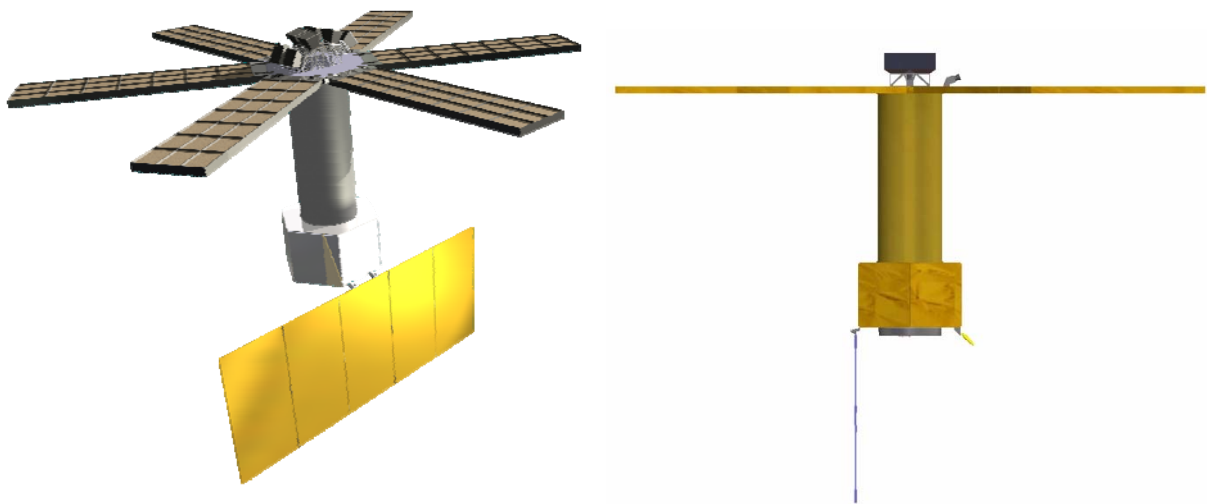

Figure 1 Schematic views of the LOFT spacecraft showing the deployed Large Area Detector (LAD) panels attached to the optical bench. The Wide Field Monitor (WFM) is placed on the optical bench at the top. The viewing direction of the LAD is co-aligned with the direction of maximum response of the WFM. In the right hand panel is shown an edge-on view of the spacecraft

# 2. LOFT SCIENCE OBJECTIVES

The main objectives of the LOFT mission are to study the effects of strong gravity and the properties of ultra-dense matter. This will be accomplished by the combined capabilities of the two instruments: the Large Area Detector (LAD) and the Wide Field Monitor (WFM).

## 2.1 LAD Science Objectives

The LAD is designed to study the neutron star structure and equation of state of ultra-dense matter and to explore the conditions of strong-field gravity by timing studies.

The neutron star structure and equation of state of ultra-dense matter is studied by measuring the mass and radius of neutron stars with high precision (<4%) by several timing related methods.

The conditions of strong-field gravity are explored by measuring the mass and spin of black holes, and by detecting general relativistic precession and the Quasi-Periodic-Oscillations introduced by matter falling into the black hole. The effects of general relativity close to the event horizon of black holes are also studied by observing the distortion of the Fe line, thus demanding a high spectral resolution of the instrument.

Common for these timing studies is the fact that the coherence time of the phenomena is short, which means that an observation with a large area detector cannot be substituted by a longer observation with a smaller detector. Also, many of the phenomena are associated with transient sources or special states of known sources, which makes a wide field monitoring instrument indispensable in achieving mission success.

## 2.2 WFM Science Objectives

The main goal for the WFM is to detect new transients and state changes of known sources suitable for observation with the LAD in order to achieve the top level goals of probing strong gravity and study the equation of state of neutron star matter. However, the WFM will be able to do important science on its own.

With a very large field of view, covering more than 1/3 of the sky, the WFM offers a high duty cycle compared to other X-ray monitors with a scanning mode of operation, like the past RXTE/ASM and the current MAXI monitor. Therefore the WFM will be particularly well suited for detecting a large number of short duration events, like Type-I X-ray bursts, Gamma Ray Bursts (GRBs), Soft Gamma Repeater (SGR) events, and also Terrestial Gamma Flashes (TGFs) and study these events with a 300 eV resolution down to 2 keV.

The source location accuracy and energy resolution are well suited for the investigation of some of the open issues in the study of GRBs. Some of the issues include: the models for the physics of prompt emission, the existence and properties of spectral absorption features and the detection of high-Z GRBs.

Another area where the WFM will contribute is in the topic of X-ray flashes. The WFM studies of the population and properties of X-ray flashes found to accompany supernova shock break-out and the disruption of stars and planetary objects by super-massive black holes should provide important results.

With a sensitivity of better than 5 mCrab in one day for Galactic Center observations, the WFM will monitor the long term variability a large number of X-ray sources.

In addition, a real time burst alert system, providing ~1' locations to ground based users in <30 s of fast transients, like gamma ray bursts, is a unique capability beyond the role of the WFM as a support instrument for the LAD.

# 3. LOFT WIDE FIELD MONITOR REQUIREMENTS

The LAD will consist of several deployable panels attached to an optical bench at the top of the space craft. The wide field monitor is placed on the optical bench, as indicated in Figure 1, showing the schematics of the LOFT space craft design. The potential field of view of the WFM is thus limited to the hemisphere centered on the pointing direction of the LAD.

In order to meet the top level science goals of the LOFT mission [1][2][4], a set of requirements has been defined for the WFM. These requirements, as well as the performance goals, are summarized in Table 1.

Table 1 Summary the WFM requirements and goals

| Item | Requirement | Goal |
|---|---|---|
| Location accuracy | <1 arcmin | <0.5 arcmin |
| Angular resolution | <5 arcmin | <3 arcmin |
| Peak sensitivity in LAD pointing direction (count rate: 5 σ) | 1 Crab (1 s)  (2-30 keV range) 5 mCrab (50 ks) | 0.2 Crab (1s)  (2-30 keV range) 2 mCrab (50 ks) |
| Absolute flux calibration accuracy | 20 % (5-12 keV range) | 15 % (5-12 keV range) |
| Relative flux calibration accuracy over Field of View | 5% (5-12 keV range) | 2.5% (5-12 keV range) |
| Relative flux calibration accuracy over time | 10% (5-12 keV range) | 5% (5-12 keV range) |
| Field of view | 1 π steradian around the LAD pointing | 1.5 π steradian, for large Sun angles part of the LAD accessible sky would otherwise not be monitored |
| Energy range | 2 – 30 keV (primary) 30-80 keV (extended) | 1.5 – 30 keV (primary) 30-80 keV (extended) |
| Energy resolution | 500 eV | 300 eV |
| Energy scale knowledge | 4% | 1% |
| Number of energy bands for compressed images | >=8 | >=16 |
| Time resolution | 300 sec for images 10 μsec for event data | 150 sec for images 5 μsec for event data |
| Absolute time calibration | 1 μsec | 1 μsec |
| Duration for rate triggers | 0.1 sec – 100 sec | 1 msec - 100 sec |
| Rate meter data | 16 msec | 8 msec |
| Availability of triggered WFM data | 3 hours (two revolutions) | 1.5 hours (1 revolution) |
| On-board memory | 5 min @ 100 Crab | 10 min @ 100 Crab |
| Broadcast of trigger time and position available to end users | < 30 sec after the event for 65% of the events | < 20 sec after the event for 65% of the events |

The WFM is based on the coded mask principle, as employed successfully in different energy bands on several previous missions like, GRANAT/SIGMA, BeppoSAX/WFC, INTEGRAL/JEM-X/IBIS/SPI, RXTE/ASM, SWIFT/BAT, and SuperAGILE. The mask shadow-gram is recorded by a position sensitive detector, and can then be deconvolved into a sky image. The size of the point spread function in the sky image is determined by the ratio of the mask pixel size and the mask-to-detector distance. The mask pixel size must always be larger than the corresponding detector resolution. The above examples include instruments, where the detector has a regular, 2 dimensional position resolution (for example pixilated detectors), as well as strictly 1-dimensional detectors, which then only produce 1-dimensional sky images. In the 1D case, two orthogonally oriented instruments or a rotating instrument can then provide accurate source positions in two dimensions.

The LOFT WFM will employ detectors providing 2D position information, but with a resolution in one coordinate, which is 50-100 better than in the other coordinate (in the project often referred to as a 1.5D detector). The design is then based on two orthogonal cameras to form a unit providing a combined fine 2D position resolution. The total WFM instrument will consist of 10 cameras forming 5 such camera units. Each camera and combined unit will cover a 45°×45° area of the sky. The overall WFM configuration of the five units and the resulting sky coverage is described in Section 5.

The imaging capabilities of the WFM are also needed to resolve any source confusion in the ~1° field of view of the non-imaging LAD instrument. Therefore the WFM will have its peak sensitivity in the LAD pointing direction.

The WFM characteristics are summarized in Table 2.

Table 2 Summary of the main WFM characteristics

| WFM Instrument Characteristic | |
|---|---|
| Detector type | Si Drift |
| Mass (10 cameras forming 5 units) | 79 kg, incl. margins |
| Peak Power (10 cameras forming 5 units) | 109 W |
| Detector Operating Temperature | <-20 °C |
| Total Detector Effective Area (5 camera unis) | 1820 cm$^2$ |
| Energy range [keV] | 2-50 keV (50-80 keV, extended) |
| Energy resolution [FWHM] | <500 eV @ 6 keV |
| Mask pixel size | 250 μm x 16 mm |
| Field of View | 180° x 90° FWZR  plus 90° x 90° towards anti-Sun  hemisphere |
| Angular Resolution (geometric) | <5 arcmin (5 arcmin × 5° per camera) |
| Typical/Max data rate after binning and compression | 50/100 kbits/s |

## 4. THE WFM DETECTORS AND CAMERA

This section summarizes the properties of the individual WFM camera.

### 4.1 The WFM Silicon Drift Detector and Read-Out

The Silicon Drift Detectors (SDDs) to be used for the WFM have the same design and characteristics as those for the LAD [4], with the only difference in the smaller overall size of the Si tile (for the WFM: 77.4 mm x 72.5 mm, vs 120.8 mm x 72.5 for the LAD) and in the smaller anode pitch (145 μm for the WFM, versus 970 μm for the LAD), and share the heritage from the ALICE particle experiment in the LHC at CERN [5][6]. The individual SDD for the WFM have the characteristics listed in Table 3. The SDDs are expected to meet the performance goal of a 300 eV spectral resolution.

The key difference between the LAD and the WFM requirements to the detector is the WFM need for position information of the X-ray interactions in order to function as a coded mask instrument. The working principle of the SDD is shown in Figure 2.

The size of the charge cloud resulting from an X-ray interaction in the Si will depend on the drift length before reaching the anodes. By fitting the charge cloud it is possible to derive three parameters about the incoming photon: (X,Y,E)=(position in the anode direction, drift length, energy). It will be possible to achieve a position resolution of ~70 μm (FWHM) in the anode direction and of 3 mm (FWHM) at 6 keV and 8 mm at 2 keV in the drift direction. The imaging performance of the SDD is described in further details in [7] and references therein.

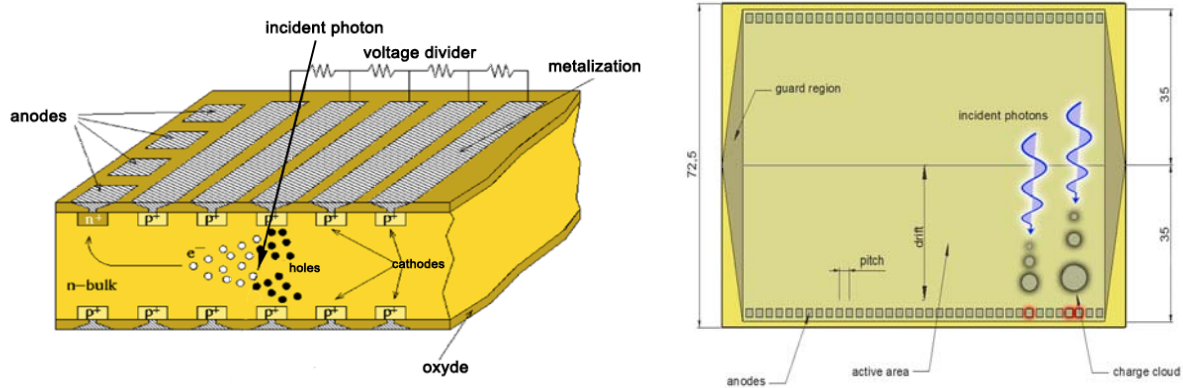

Figure 2 Illustrations of the working principle of the Silicon drift detectors. The absorption of a photon in the Silicon bulk is shown to the left. To the right is shown the detector module with read-out anodes at the top and the bottom. The size of the charge cloud is a function of the drift length and is fitted to provide the position in the drift direction

Table 3 Summary of the characteristics of one SDD detector module. One WFM camera will contain 4 detector modules

| Parameter | Value |
|---|---|
| Si thickness | 450 µm |
| Si tile geometric size | 77.4 mm x 72.5 mm |
| Si tile active area | 65.1 mm x 70.0 mm = 45.57 cm$^2$ |
| Anode pitch | 145 µm |
| Number of read-out anodes per tile | 448 x 2 rows = 896 total |
| Drift length | 35 mm maximum |

A total of 4 SDDs compose the detector plane of each individual WFM camera. The overall dimension of the SDD assembly is 145 mm x 154.8 mm. The active area of each WFM camera is a square of 142.5 mm x 142.5 mm. This allows arranging two identical cameras with a 90° relative rotation (in order to achieve fine angular resolution in two coordinates) but still having the same field of view, to compose one WFM unit.

The choice of the SDD size is driven by the requirement of a square active area for the overall camera, and by the requirement of not increasing the drift length longer than 35 mm. The choice of the anode pitch is the result of an optimization study of the detector performance, based on experimental tests and by simulations [7].

The read-out of the anode signals in one detector module is performed by 2 rows of 14 ASICs mounted on the front end electronics board, which is mounted on the back side of the Si detector, as illustrated in Figure 3. The maximum effective area of a detector module is ~45 cm2, and is shown as function of energy in Figure 4.

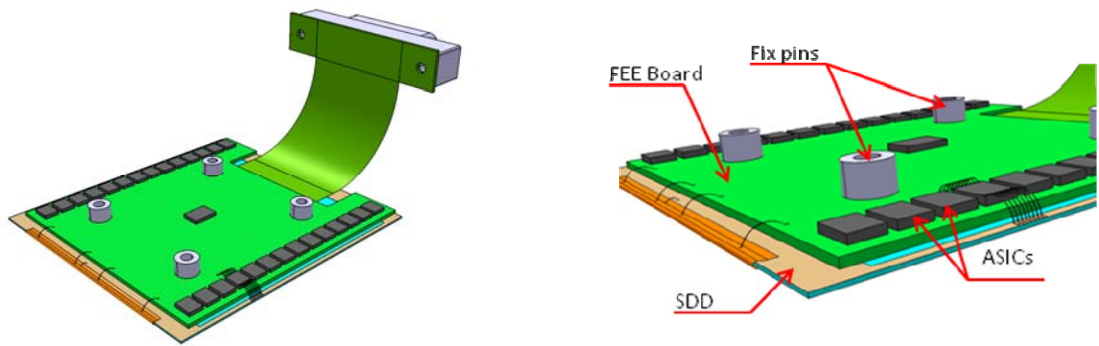

Figure 3 A view of the front end electronics (FEE) board and the ASICs mounted on the back side of one of the four SDD detector modules in one WFM camera

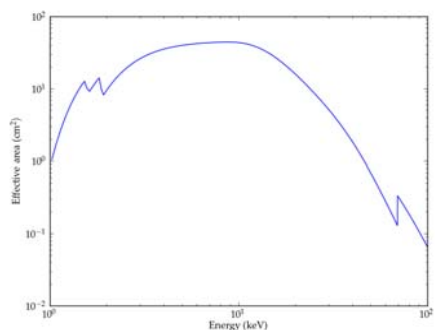

Figure 4 The effective area, as function of energy, of one WFM detector module with a maximum of ~45 cm$^2$. As the WFM coded mask has an open fraction of 25% this figure (at least for energies below 30 keV) also corresponds to the on-axis effective area of one WFM camera, which contains 4 detector modules

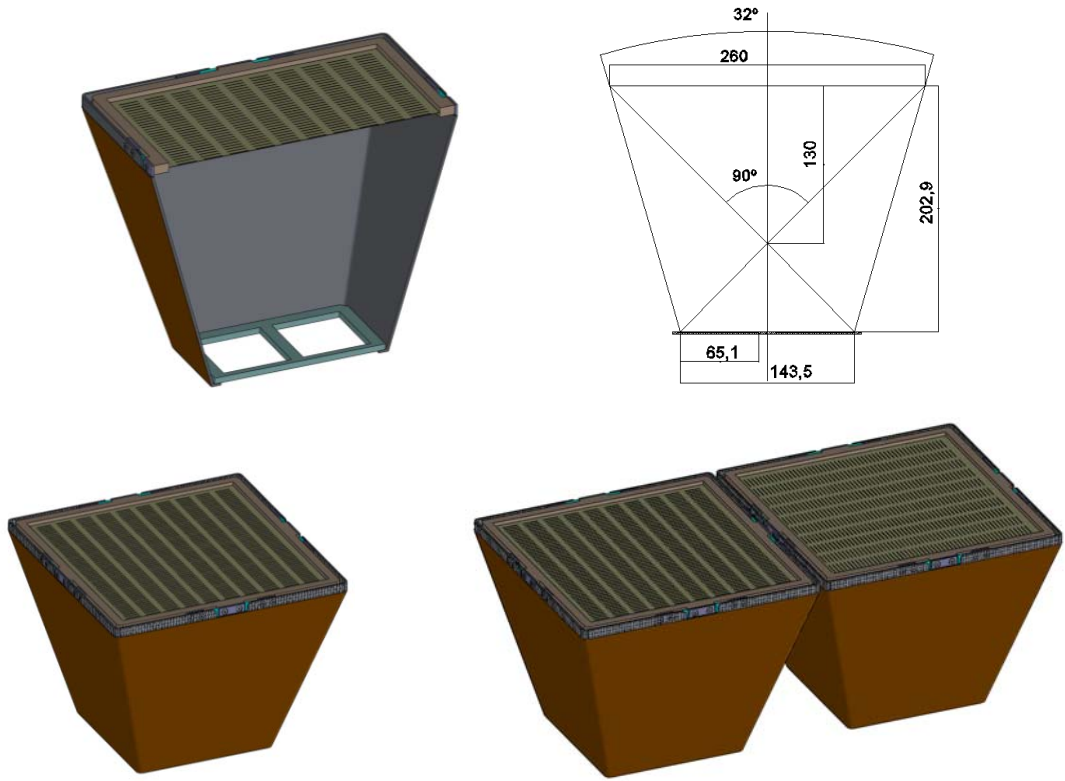

Figure 5 View of half of one WFM camera (upper left), indicating the detector tray at the bottom, the collimator, and the coded mask at the top. The dimensions and the viewing angles (full and zero illumination) are indicated (upper right) with units in mm. At the bottom is shown one camera (lower left) and one camera unit consisting two orthogonally oriented cameras (lower right)

### 4.2 The WFM Camera

The WFM camera square detector plane consists of 4 SDD detector modules mounted in a detector tray offering a maximum effective detector area of ~180cm$^2$ for one camera (see Figure 5).

A coded mask made of Tungsten (W), with a size of 260x260 mm, will be placed at 203 mm above the detector plane. The thickness of the mask is 150 μm to ensure opaqueness and coding in the primary energy range up to 30 keV. The size of the mask elements is 250μm × 16 mm, thus ensuring the position resolution of the detector to oversample the mask pixels by a factor of at least two at all energies within the nominal range [8], and providing a geometrical resolution better than 5 arcmin in the anode direction of fine position resolution. A thermal shield of 7.62 μm Kapton foil will be placed on top of the mask.

The open fraction of the coded mask is chosen to be 25%. Thus, the on-axis effective area corresponds to a maximum of 45 cm2 (see Figure 4) for one camera. The rationale behind the choice of a reduced open fraction, compared to the often used 50%, is two-fold: First, it reduces the event rate from X-ray sources and the cosmic diffuse X-ray background (CXB) by a factor of two, reducing the telemetry requirements. Secondly, as the background is dominated by the CXB [9] (and other sources), the signal to noise of fitting weaker sources is actually improved relative to a 50% open mask [10]. The "cost" of this choice of open fraction is a reduced signal to noise for sources brighter than the background. The mask pattern is currently based on a bi-quadratic residue set based on the prime number 16901. This set has perfect coding properties for an ideal detector, i.e. a PSF with perfectly flat side lobes, although a random mask may also be considered.

The collimator supporting the coded mask is made of 2 mm Carbon-fiber-reinforced polymer (CRFP) to provide mechanical stability, which is essential for the performance of the fine pitch mask. The collimator is covered by a layer of 150μm thick Tungsten shield preventing X-rays from outside the field of view from reaching the detector. In addition, thin layers of Cu and/or Mo may also be introduced, as simulations show that fluorescence lines from these materials induced by the hard component of the CXB may be used for energy calibration of the detector in space, thus avoiding the use of onboard radioactive sources (See [9] for further details).

The opening angle of the collimator defines the zero response field of view along the detector axes to be 45° off-axis, or a 90° × 90° square (see Figure 5, upper right panel). The field of view, where the detector is fully illuminated by a source corresponds to a 32° × 32° field (sometimes for coded mask instruments referred to as the "fully coded field of view"). The off-axis response is determined by several factors: the cosΘ projection factor, partial illumination of the detector due to the collimator for angles larger than 16°, and mask vignetting, which is caused by the finite mask thickness to mask pitch ratio. The mask vignetting only plays a role in the direction of fine mask pitch.

Studies have shown that with the large viewing angle of the WFM camera, the impacts of micro-meteorites and small particles of orbital debris will pose a risk for the Si detectors. To mitigate this, a thin (25 μm) Beryllium filter will be placed ~8 mm above the detector plane. In combination with the thermal blanket in front of the coded mask, this Be filter will reduce the risk of impact a particle of size ~100 μm to about $1.6 \times 10^{-2}$ per year per camera. The low energy response is mildly affected by the Be filter, which has a transparency of ~70% at 2 keV.

### 4.3 The WFM Camera Unit

Each SDD detector has a fine (~30-70 μm, energy dependent) position resolution in one direction and much coarser (~5-8 mm) in the other direction. This is reflected in the asymmetrical design of the coded mask elements, providing each camera with an angular resolution of ~5 arcmin × ~5°. The fine position resolution in the two coordinates is guaranteed by 2 orthogonal and co-aligned cameras forming a WFM camera unit (Figure 5, lower right panel).

The imaging properties of the two crossed cameras have been studied in detail and shown to live up to the performance requirements. An example of a part of a 1.5D image of the Galactic center region is shown in Figure 6. For further details, see [11].

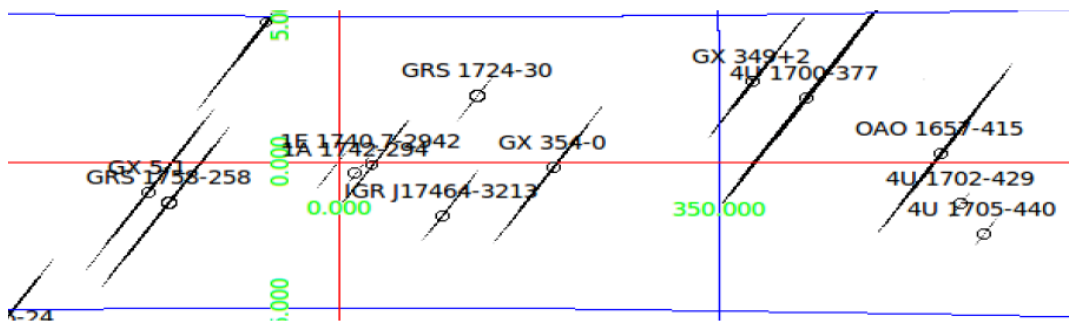

Figure 6 An approximately 30° x 10° part of an image derived from a simulated 10 ks observation of the Galactic center region with one WFM camera (shown in galactic coordinates) to illustrate the shape of the elongated WFM PSF with a size of ~5' x ~5° (adapted from [11])

## 5. WFM CONFIGURATION AND SKY COVERAGE

The WFM must cover as large a fraction of the sky, accessible to the LAD, as possible in order to serve its purpose as a monitoring instrument. The LAD nominal pointing directions are, mainly for thermal reasons, confined to a band on the sky of ±30° width, perpendicular to the Sun direction. This corresponds to an area of 50% of the sky. In addition, the LAD is expected to be able to temporarily point in the cap around the anti-Sun direction, thus being able to access 75% of the sky. In order for the WFM to be able to detect sources for LAD follow-up, the field of view needs to be large, still maintaining a good sensitivity, within reasonable resources of mass, power and telemetry. At the same time the WFM configuration should be easily accommodated on the spacecraft, thus for example, making the placement of a WFM unit pointing in the opposite direction of the LAD pointing less practical.

The WFM will be placed on the optical bench, which is orthogonal to the LAD pointing direction. Trade-off studies have converged towards a WFM configuration consisting of 5 units (a total of 10 cameras) each with the size and field of view

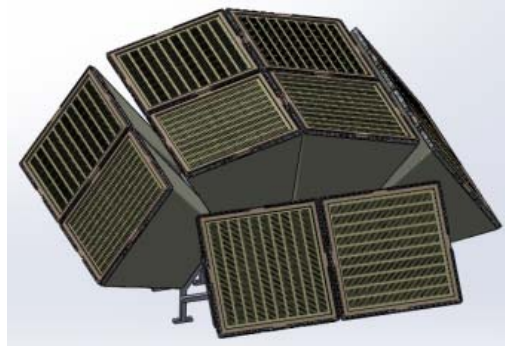

Figure 7 The overall logical arrangement of the 5 WFM camera units. 4 units are arranged in a half-circle centered around the LAD pointing direction (at −60°, at −15°, at +15°, at +60°), while the 5th unit is placed at 60° from the LAD direction towards the anti-Sun direction

described in Section 4. The WFM baseline configuration is illustrated in Figure 7. This configuration will cover ~5.5 steradian or ~44% of the sky at zero response, and ~4.2 steradian or 1/3 of the sky at 20% response relative to on-axis.

Four of the five units are arranged such that their viewing axes lies in the plane of the solar panel of the LOFT spacecraft, and the fifth unit is tilted out of this plane, away from the Sun, by 60˚. The viewing directions of the four units are off-set by ±15˚ and ±60˚ relative to the LAD pointing direction, which also lies in the solar panel plane (see Figure 1). The effective area of the full WFM assembly is shown in the right hand panel of

Figure 8. With this arrangement, the two central units have the LAD target in their field, where the detectors are fully illuminated, providing the maximum WFM coverage, with ~160 cm$^2$ effective area, in the direction observed by the LAD. The zero response field of view of the 4 units really extends to ±105˚ × ±45˚(210˚ × 90˚). However, depending on the configuration of the LAD panels and the placement of the WFM units on the optical bench, only an unobstructed field of view of 180˚ × 90˚ can be assured. The 60° tilt of the two outer units with respect to the LAD pointing direction is preferred in order to have a reasonable response at the edge defined by the plane of the optical bench. In this configuration, the WFM nominally covers half of the sky that is accessible to LAD pointings. The WFM may therefore in 2 LOFT pointings cover all the sky accessible to the LAD, or an area corresponding to at least 75% of the sky.

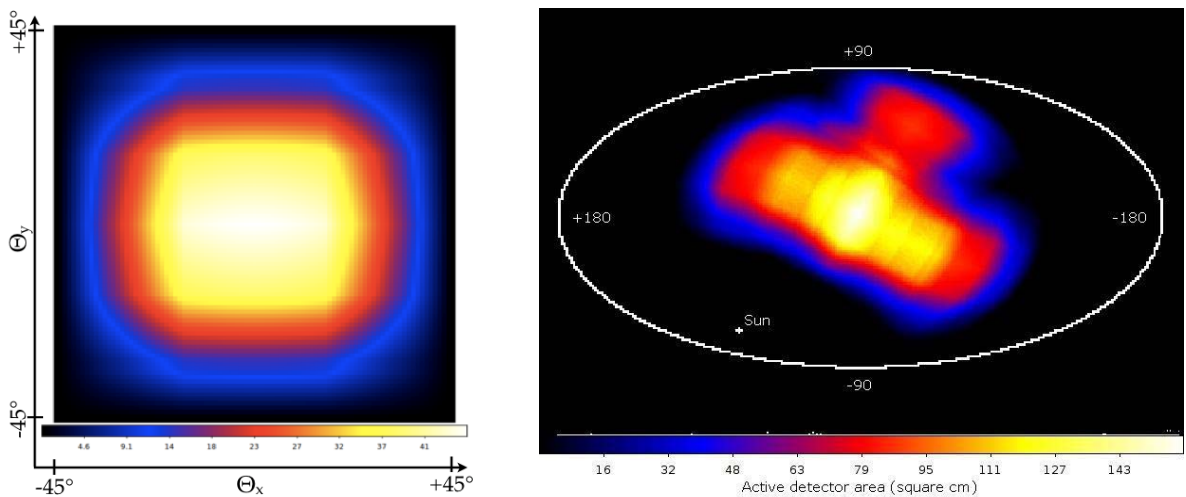

Figure 8 Left panel: map of the single camera sensitive area expressed in cm$^2$, with a maximum of ~45 cm$^2$ of a single camera. The map takes into account the main geometrical effects (mask open fraction, vignetting, shadowing of the collimator walls, detector non-sensitive areas, cos(θ) effect). Right panel: map in Galactic coordinates of the active detector area for an example observation of the Galactic center. The effective area has its maximum of ~160 cm$^2$ in the direction of the LAD pointing. The figure does not include any obscuration at the edge of the FoV by the optical bench.

We note that the WFM configuration has several overlapping fields of view between different units. This will provide a good handle of the in-flight calibration of the off-axis response, as sources will be simultaneously observed at different off-axis angles in individual cameras. In particular, a WFM unit, with two orthogonal cameras pointing in the same direction, will provide excellent opportunities to calibrate the mask vignetting. This self-calibration property of the configuration will allow reducing the systematics in flux determination, which will be essential for long term monitoring of variations in X-ray sources. In order to further reducing the systematics in the overlapping fields of view, and avoiding source confusion, the camera units will be rotated relative to each other by small angles (<±10°). These rotations are not shown in Figure 7, which only illustrates the general sky coverage.

A stable thermal environment is essential for the detector performance, but also for the mechanical stability of the collimator and mask assembly. Therefore a sunshade will be placed on the optical bench to prevent direct illumination of the WFM masks by the Sun during normal observations.

The WFM instrument performances are specified in Table 4.

Table 4 The WFM expected instrument performance specifications based on the described SDD detectors

| Parameter | One Camera | One Unit = 2 crossed cameras | Overall WFM |
|---|---|---|---|
| Energy Range | 2-80 keV (ext) | 2-80 keV (ext) | 2-80 keV (ext) |
| Active Detector Area | 182 cm$^2$ | 364 cm$^2$ | 1820 cm$^2$ |
| Peak Effective Area (on-axis, through mask) | >40 cm$^2$ | >80 cm$^2$ | >80 cm$^2$ |
| Energy Resolution FWHM | < 300 eV EOL @ -30°C | < 300 eV EOL @ -30°C | < 300 eV EOL @ -30°C |
| Field of View at Zero Response | 90° x 90° | 90° x 90° | 210°x90°+45°x90° |
| Field of View at 20% response | 60° x 60° | 60° x 60° | 180° x 60°+60° x 60° |
| Angular Resolution | 5' x 5° | 5' x 5' | 5' x 5' |
| Point Source Location Accuracy (10σ) | < 1'x30' | < 1'x1' | < 1'x1' |
| On-axis sensitivity at 5σ in 3 s (Gal. Center) | 380 mCrab | 270 mCrab | 270 mCrab |
| On-axis sensitivity at 5σ in 58 ks (1 day Galactic Center pointing) | 3.0 mCrab | 2.1 mCrab | 2.1 mCrab |

## 6. WFM DATA FLOW, TELEMETRY AND DATA MODES

This section provides an overview of the WFM data flow and the limitation imposed by the telemetry constraints.

### 6.1 WFM digital electronics and data flow

The WFM digital electronics and overall data flow is illustrated in Figure 9. The signals from the ASICs in each of the 5 camera units are treated in the 5 back end electronics (BEE) units located in a common box together with the 5 power supply (PS) units. The main task of the BEE is to process the anode signals when the ASIC is triggered by an interaction in the detector. Unwanted triggers from particle interactions are filtered off. By fitting the Gaussian shape of the charge cloud signals after proper pedestal and common noise subtraction, the position and energy of the incoming photon is determined. The X-ray event data (X,Y,E,T) also includes the time of the interaction with an accuracy of ~5μs. This event data packet can by applying differential time tagging be transmitted in 40 bits per event.

In science mode, the X-ray event data generated by the BEEs constitute the basic data to be further processed by the central data handling unit (DHU). The expected data rate for one WFM unit is ~550 c/s for the CXB background, while the Crab count rate will be ~275 c/s. In order to facilitate onboard localization of burst sources the DHU is supplemented by a board with hardware optimized for the discrete FFT calculations required for image deconvolution.

### 6.2 Diagnostic and electronic calibration modes

In order to verify the detector performance and calibrate the parameter settings a diagnostic mode will provide the full information of the ASIC readouts. An electronic calibration mode will verify the gain by charge injection into the SDDs.

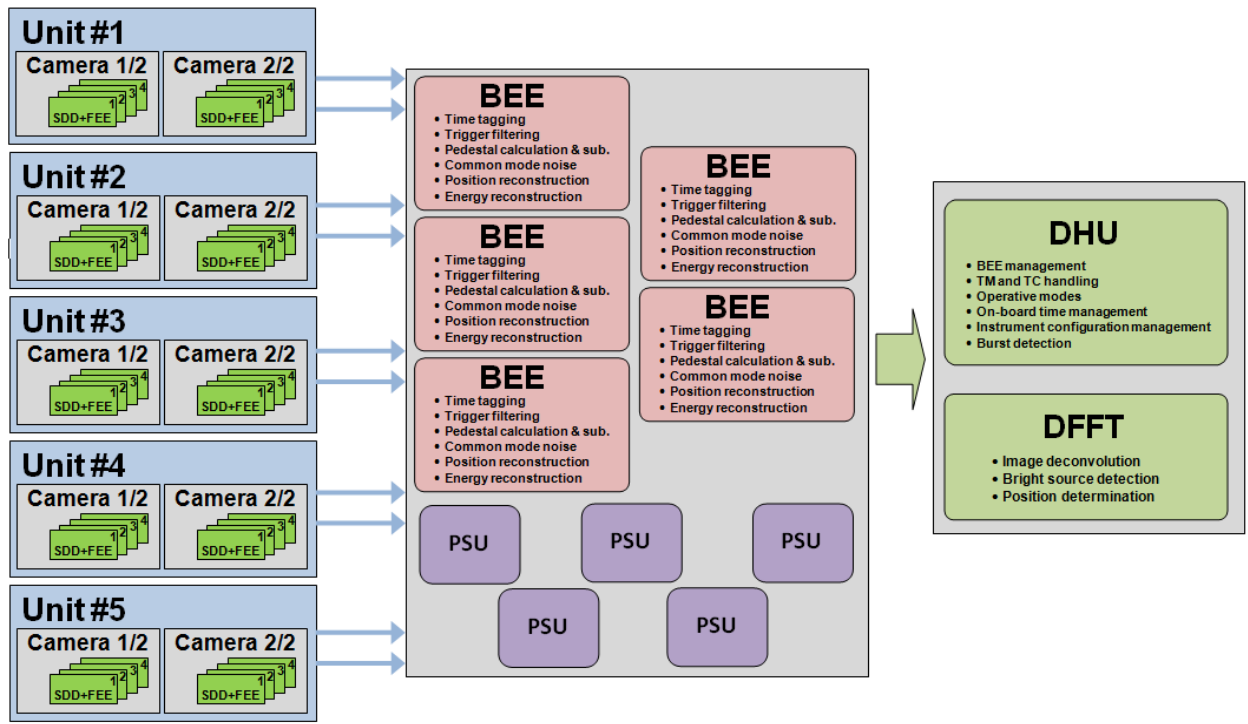

Figure 9 The WFM data flow from the 5 camera units to the 5 back end electronics (BEE) and to the Data Handling Unit (DHU), which in turn is connected the spacecraft onboard data handling system. Each BEE and 2 cameras are powered by a separate power supply unit (PSU). With the DHU is also shown the source localization functionality provided by the discrete fast Fourier transform (DFFT) hardware.

### 6.3 Normal data taking

The maximum, sustained data rate for the WFM will normally be limited by the down-link bandwidth to ~100 kbits/s averaged over the orbit. Therefore it will not be possible to downlink the full event-by-event information and some data binning is required. The data rate for uncompressed event-by-event data is given for one camera unit as function of the flux of the sources in the field of view in Table 5. We note that for 5 units observing blank fields with only the CXB the telemetry budget will already be exceeded.

The normal data mode will consist of 3 binned data products:

- Detector images integrated over 300s in 8 or 16 energy bands
- Full resolution detector spectra integrated over 30s
- Detector rate-meter data with 16 ms resolution in 8 energy bands

The bulk of these data is the detector images. The images are for typical data rates and 8 or 16 energy bands sparsely filled (much less than 1 count per pixel). This allows for efficient compression by transmitting the distance between filled pixels. The compressed data rate of detector images for one unit as function of source flux is shown in Figure 10. The optimum number of bits per event used in the encoding is easily determined by the count rate, We note that the compressed data rate for detector images for one unit observing sources corresponding to 10 Crab on-axis will be less than 13 kbits/s for 16 energy bands integrated for 300 s. As the typical rate in each of the 5 units is normally much lower, the image data with these parameters will fit well into the telemetry budget. The highest rates are of course expected from units observing the Galactic Center region with several bright sources. Depending on the final observing strategies, the parameters for image integration time and/or the number of energy bands may be adjusted dynamically or by command in order to optimize the telemetry usage. We in particular note, that (as illustrated in Figure 10) the difference in telemetry rate for 8 and 16 energy bands is not dramatic. This is explained by the simple fact that the distance between filled pixels will double, when switching from 8 to 16 energy bands, requiring only 1 more bit per event to encode the distance between filled pixels.

Table 5 The average data rates for full event-by-event mode and for triggered data for one camera pair to be compared with the normal averaged telemetry allocation of <100 kbits/s. Note that for triggered data the source flux is the average over the 300s to be transmitted and not the peak flux of the transient event

| Total rate in 1 unit (2 cameras) | Data rate for 40 bit event format | Average data rate over 1 orbit to transmit 300 s of event-by-event data |
|---|---|---|
| CXB | 22 kbits/s | 1.2 kbits/s |
| CXB + 1 Crab source | 31 kbits/s | 1.7 kbits/s |
| CXB + 10 Crab source | 113 kbits/s | 6.3 kbits/s |
| CXB + 30 Crab source | 295 kbits/s | 16.4 kbits/s |
| CXB + 100 Crab source | 932 kbits/s | 51.8 kbits/s |

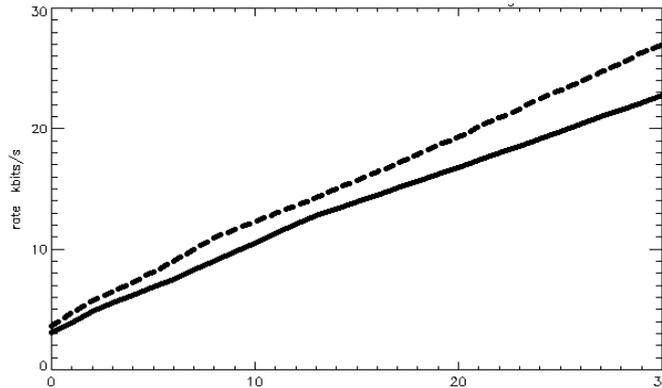

Figure 10 The compressed data rate for the detector images (in kbits/s) for one WFM unit (2 cameras) as function of the total flux from point source in the field of view in units of the Crab flux for 8 energy bands (solid line) and 16 energy bands (dashed line) for an integration time of 300 s (CXB corresponds to ~550 c/s and 1 Crab to ~275 c/s)

### 6.4 Event-by-event data and trigger mode

X-ray events of duration significantly shorter than the normal integration time for detector images of 300 s will be difficult to identify in the normal date. Therefore, a burst trigger logic will operate to detect increases in flux on time scales from a fraction of a second to ~100s. The burst trigger logic is active during data taking mode and is based on detecting increases in the count rate of X-ray events in the WFM cameras, which are likely to originate from activity in cosmic X-ray sources and not from increases in the general background. When the onboard software detects a potential transient event it will save data for transmission in the "event-by-event" format in order for the ground software to be able to analyze the event in full detail. Normally a burst trigger will mark an interval of data of 300 s around the trigger time to be transmitted in event-by-event format. The estimated 'cost' in telemetry per trigger averaged over the orbit is listed in Table 5 for some average flux values during the 300 s interval. These numbers may then be compared to the overall telemetry budget of <100 kbits/s.

The types of sources expected to trigger the event-by-event mode will be Type I and Type II X-ray bursts, soft gamma repeaters, gamma ray burst and X-ray flashes, as well as terrestrial gamma-ray flashes. Some highly variable sources may also trigger the system. In order to avoid transmission of event-by-event data from such 'uninteresting' sources or by background induced false triggers, the burst trigger logic will make use of an onboard localization algorithm to identify the position of the source (see Section 7) and filter triggers based on entries in an onboard catalog of X-ray sources.

### 6.5 The WFM on ground data processing

The LOFT ground segment will include a Science Data Center (SDC). For the SDC a high priority task will be to analyze the WFM data immediately, or as soon as possible, after the reception of the science data in order to identify new sources and state changes in known sources, that will warrant a target of opportunity for the LAD. The quick-look analysis results from the WFM are also expected to serve as input for target-of-opportunity for other space or ground based observatories.

# 7. THE LOFT BURST ALERT SYSTEM

It is expected that the WFM will detect ~150 Gamma Ray Bursts per year. With a response down to 2 keV and a good energy resolution the WFM will offer unique capabilities in this field of research in its own right.

Scientifically it is highly desirable to observe these sources with other telescopes and instruments as soon as possible after (or even during) the event. Therefore LOFT will employ a VHF transmission capability to send a short message about the occurrence of such events with minimum delay to a network of VHF receiving stations on the ground for further distribution to interested observatories. The LOFT burst alert system will distribute the location of a transient event with ~1 arc minute accuracy to end users within 30 s (goal 20s) of the onboard detection of the burst in at least 2/3 of the cases.

## 7.1 Onboard Source Localization

The onboard software will localize the position in the sky of the source responsible for the burst trigger. For a coded mask instrument, the deconvolution of the detector image into a sky image is computationally very demanding. The deconvolution will be done onboard by cross correlation of the mask pattern with the background subtracted detector image. This is most efficiently done by a discrete Fast Fourier Transform (DFFT) algorithm. As a result of burst trigger, the onboard software will enter a mode to determine the position of the source. The DFFT transformation from coded mask detector images into sky images will be performed by a special board with hardware optimized for performing the DFFT.

If the rate increase can be localized as a point source, and thereby confirmed to be an outburst of a real X-ray source, the position is initially defined relative to the camera coordinate system. The position is then, based on the pointing information, transformed into a position on the sky, which is compared with a catalog of known X-ray sources. If the position does not correspond to a known source and the calculations meet a certain set of quality/reliability criteria the software will send a short message with brief information about the event to the onboard data handling (OBDH) in order for it to be transmitted immediately to the ground via the spacecraft VHF transmitter system. The message will contain information on burst time, burst location, duration, and a set of quality flags for the use of the ground based users. The total amount of data to be transmitted is on the order of 1 kbits. The "event-by-event" data that form the basis of the onboard localization will be stored for transmission through the normal telemetry channel during the next regular ground station pass, as described in Section 6.4. Off-line analysis will validate and improve the location error box and other information distributed in the original burst alert packet

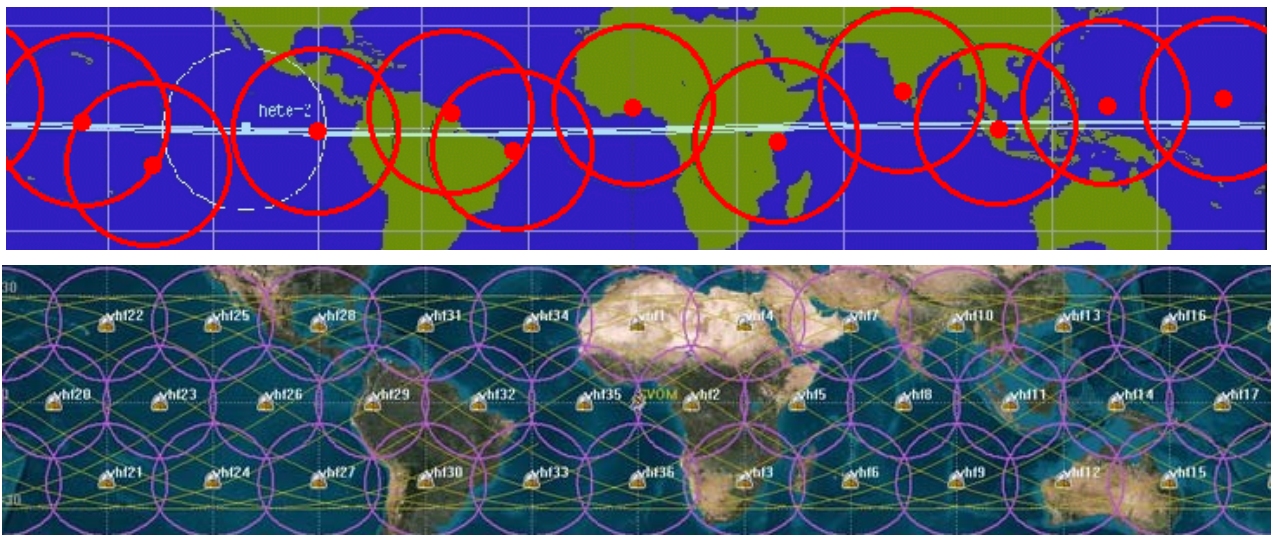

Figure 11 The location of VHF stations for the former GRB mission HETE-II (top) and the theoretical SVOM network (bottom). The ideal LOFT burst alert system network consists of the 12 central/equatorial stations of the SVOM system

## 7.2 Burst Alert System Ground Segment

The LOFT burst alert system ground segment is going to be based on the network of VHF stations planned for the French-Chinese mission SVOM. The typical data rate for this system is 600 bits/s, which allows a short message containing the basic burst information, to be transmitted in less than 2 seconds. In defining the requirements of a maximum delay between the burst trigger and the delivery of the burst information packet of <30s a very conservative approach has been taken in calculating the worst case delay introduced by each element in delivery the chain. A realistic goal will be a <20 s delay, with the system in many cases performing significantly better.

With LOFT in a close to equatorial, ~550-600 km altitude orbit, the number of VHF ground stations needed to ensure continuous coverage can in principle be limited to 12 ideally placed stations, as opposed to the 36 needed for the higher inclination SVOM mission. However, the placement of the stations will eventually be determined by available land based sites with sufficient infrastructure (see Figure 11), as was the case for the HETE-II mission.

The VHF ground station will be managed by a central LOFT Alert Center, having the responsibility of validating the burst alerts and distributing the alerts to the end users, for example through the GCN network commonly used for sending out GRB alerts.

We note that the LOFT mission baseline does not include any capability for the satellite to do automated reorientations to observe the GRB afterglows with the LAD, as this would impose significant and costly requirements on the spacecraft autonomy.

## 8. CONCLUSIONS

The LOFT mission promises to significantly advance our knowledge about some of the fundamental questions related to strong gravity and dense nuclear matter. In addition, and not least, it will serve as a general X-ray observatory to study many classes of X-ray sources. The WFM instrument will play a significant role in this, partly as a supporting instrument, but also as a result of its superior monitoring capabilities. In particular, the WFM is expected to provide significant, independent contributions to the field of the gamma ray burst studies through the near real time burst alert system.

According to the current ESA down selection schedule, it will be known by early 2014, if LOFT will move ahead from the assessment phase and be implemented for an expected 2024 launch. In the unfortunate event that the LOFT mission will not be selected, the WFM instrument design will, however, form a solid base for inclusion in other mission concepts.